# Adaptive solitons

Yuchuan Chen[1]*, Jiacheng Zhao[2,3], Qiuran Zhu[2], Bill Chen[1]

[1]Lasoliton LLC, Andover, MA 01810

[2]The Institute of Optics, University of Rochester, Rochester, NY 14627

[3]Center for Matter at Atomic Pressures, University of Rochester, Rochester, NY 14627

**Abstract:** Femtosecond laser pulses have incredible potential but still face some performance and stability issues. Here we discover a new nonlinear optical effect, frequency-comb-induced stimulated Brillouin scattering resonance, which can only be excited by a new type of soliton, adaptive solitons. Adaptive solitons possess the unique properties of adaptive down-chirp which breaks area theorem limitation on pulse energy in normal dispersion lasers, spectral profile self-regulation which provides high pulse contrast and breaks gain narrowing limitation in amplifiers, and freedom from dispersive radiation that weakens and deforms current ultrashort laser pulses in transmission media. More intriguingly, they have an adaptive transmission gain in normal dispersion waveguides which has the potential to enable them to be lossless in long normal dispersion waveguides for repeaterless soliton communication. A passive frequency-modulation mode-locked fiber laser was successfully built and generated incredibly stable adaptive solitons. Even though the laser had no negative dispersion elements, the output solitons were measured as down-chirped which enables them to keep soliton properties in normal dispersion waveguides, perfect for ultrafast applications with waveguide delivery. With the advantages of high consistency, high pulse contrast, and high energy capability; adaptive solitons are excellent for fields including micromachining, laser surgery, quantum computing, and spectroscopy.

**Keywords:** adaptive soliton, frequency-comb-induced Stimulated Brillouin Scattering Resonance (SBSR), passive frequency-modulation mode-locking, adaptive down-chirp, adaptive transmission gain, soliton communication.

## 1. Introduction

Ultrafast laser pulses are known for their broadband spectra, ultrashort time durations, and high peak powers. They drive numerous emerging applications including bio-imaging[1], laser surgery[2,3], micromachining[4-6], quantum computing[7-10], and many more[11,12]. These pulses are the combination of a wide range of frequencies, with their phases synchronized at their peaks. As such, broad spectra are critical for achieving ultrashort pulse durations and underpin techniques such as chirped pulse amplification (CPA). However, these broad spectra also introduce the persistent challenge of chromatic dispersion. Along with nonlinearities in laser pulse generation and transmission, dispersion can cause severe pulse degradation, making it difficult and expensive to compensate for the resulting distortions in many applications[13,14]. One promising solution to this challenge are solitons, strongly stable wave packets that serve as localized solutions to wave propagation equations[15-17]. These self-reinforcing pulses naturally balance dispersion and nonlinear effects, allowing them to travel through dispersive media without distortion. In practice, current ultrafast pulses still have some lingering flaws with dispersion during transmission.

Chromatic dispersion can be broken down into two categories: even-order and odd-order. Even-order dispersion causes chirp. Odd-order dispersion leads current laser pulses to dispersive (Cherenkov) radiation and complex pulse distortions[18-20]. The main even-order and odd-order dispersions are group velocity dispersion (GVD) and third-order dispersion (TOD), respectively. The only known nonlinear optical effect in transmission media that can balance dispersion is Kerr-induced self-phase modulation (SPM), which causes the laser pulse to experience an up-chirp, increasing its frequency over time. The up-chirp caused by SPM can balance down-chirping from negative GVD. This only enables solitons in anomalous dispersion waveguides with negligible TOD. In other words, there is no localized solitary solution for the nonlinear Schrödinger equation containing TOD[18-20], as well as positive GVD, because of the lack of a nonlinear optical effect to counterbalance TOD and positive GVD in transmission media.

At the beginning of the 1990s, the soliton concept was extended to solitary waves in nonlinear optical systems, where nonlinear gain or loss mechanisms are used to periodically balance nonlinear optical effects and dispersions[17,21,22]. Hence, the laser pulses in mode-locked lasers are often called as solitons, though most of them decay into stretched-pulses outside of their laser cavities. According to the dispersion balance approach used, current mode-locked lasers that generate compressible ultrafast laser pulses can be categorized into three types: dispersion-managed solitons, chirp-free solitons, and dissipative solitons. Dispersion-managed soliton lasers are designed such that the positive and negative GVD of different optical components counteract each other. The laser pulses in the cavity are alternately stretched and compressed more than one order of magnitude[23]. They are known as stretched-pulses outside of their laser cavities. Depending on the output position, they could be up-chirped or down-chirped. Almost all femtosecond free-space solid-state lasers are built as this type with adjustable dispersion to minimize the transform limit (TL)[24,25]. In contrast, chirp-free soliton lasers primarily consist of negative even-order dispersion components so that they can be balanced by SPM for chirp-free soliton generation[25,26]. Without the capability to excite a nonlinear effect to balance TOD, they suffer from dispersive radiation and will breakup at zero-GVD in transmission waveguides[18]. In laser cavities, dispersive waves, including dispersive radiation caused by TOD, will accumulate under phase matching conditions and generate Kelly sidebands to the oscillating pulse[25-27]. The interval of Kelly sidebands decreases with cavity dispersion, and thus chirp-free solitons feature several Kelly sidebands. Finally, dissipative soliton lasers introduce spectral filters and saturable absorbers (SA) into normal dispersion cavities to periodically balance SPM and positive GVD for dissipative soliton operation[28-30], where dissipative solitons are up-chirped[28]. Because spectral filters and SA prevent the accumulations of dispersive waves, dissipative solitons generally do not have Kelly sidebands. However, without locally balancing nonlinearity, dissipative solitons intrinsically have SPM that is too strong, causing distinct spectral peaks[17,28,29]. Dissipative solitons also can be generated from net anomalous dispersion lasers if the balance of SPM and negative GVD is broken by the nonlinear gain or loss effects[31]. They are essentially stretched-pulses in normal dispersion transmission media. Chirp-free and compressible dissipative solitons in their laser cavities are restricted by their own area theorems, which narrow their pulse durations as energy increases[26-29]. Meanwhile, the peak powers are limited by stimulated Raman scattering[32,33]. Therefore, it is hard for them to generate stable ultrashort pulses with large energy. Kelly sidebands and SPM peaks correspond to the distortions of incompressible picosecond pedestals and undesirable sub-peaks, significantly lowering pulse contrast (Fig. 1).

Here we discover a new nonlinear optical effect, frequency comb induced stimulated Brillouin scattering resonance (SBSR), capable of balancing both SPM and dispersions (GVD and TOD) for adaptive soliton generation and transmission. This balance eliminates Kelly-sidebands and SPM peaks, giving a high pulse contrast. With the SBSR effect, adaptive solitons in normal dispersion waveguides have a counterintuitive adaptive transmission gain (Fig. 1 and “Adaptive soliton” section). The effect has the potential to lock adaptive solitons to a stable state in a long transmission waveguide.

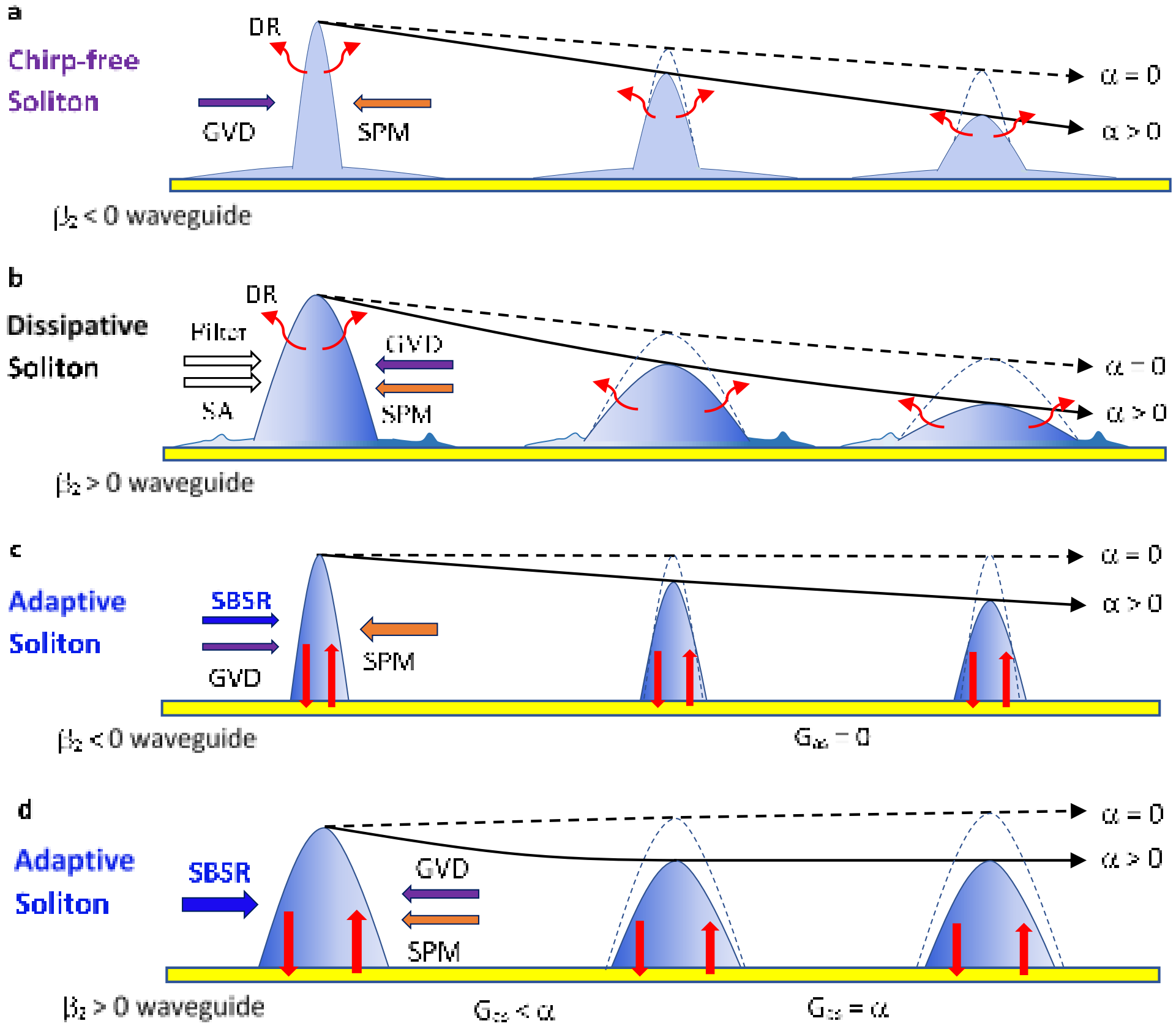

**Fig. 1 Concepts of solitons and their propagation in waveguides**

The soliton propagation direction and peak power variation are schematically shown by black dash ($\alpha = 0$) and solid ($\alpha > 0$) arrows, where α is waveguide attenuation. The horizontal arrows for Kerr-induced SPM (orange), GVD (purple), the new SBSR effect (blue), as well as filters and saturable absorbers (no fill), point to the directions of their chirp effect. Leftwards means it up-chirp the pulse, increasing the pulse frequency from the leading edge to the trailing edge (rightward is down-chirp). The gradient color in the laser pulses indicate chirp, lighter color meaning higher frequency. (a) Chirp-free solitons balance up-chirp SPM and down-chirp GVD in anomalous dispersion ($\beta_2 < 0$) waveguides. (b) Up-chirped dissipative solitons are essentially stretched-pulses outside the cavities, where there is nothing to locally balance the up-chirp effects of SPM and positive GVD in normal dispersion ($\beta_2 > 0$) waveguides. Both chirp-free and up-chirped dissipative solitons cannot fully retain energy, pulse shape, or duration during transmission, even if the waveguide attenuation were to be zero ($\alpha = 0$), due to dispersive radiation (DR, indicated by red curved arrows) caused by unbalanced TOD. Down-chirped, high-contrast adaptive solitons can be obtained both in (c) anomalous dispersion waveguides and in (d) normal dispersion waveguides where the down-chirping of SBSR effect locally balances the up-chirping of SPM and dispersions. With the SBSR effect, adaptive solitons have energy transfer (red vertical arrows in the pulse) with waveguides, absorbing phonons at the leading edge and emitting phonons at the trailing edge, which makes adaptive solitons in normal dispersion waveguides have an adaptive transmission gain ($G_{as} > 0$). They do not suffer from DR because odd-order dispersion is balanced by the SBSR effect.

## 2. Results

### 2.1 SBSR effect

An intense laser beam propagating through a medium will interact with acoustic phonons in the medium, resulting in the scattered laser experiencing either a slight frequency downshift (Stokes) or upshift (anti-Stokes), which is known as Stimulated Brillouin Scattering (SBS). While there are some applications that leverage SBS such as using backward SBS for dissipative Kerr soliton generation[34] and using forward SBS to lock a harmonic mode-locking laser to a designed harmonic order[35,36], SBS is typically undesirable and causes noise to most mode-locked fiber lasers. The mechanisms of backward and forward SBS in waveguides are well established for cases where a laser pulse is treated as multiple optical frequency lines participating in SBS interactions[34] or as an energy pulse driving the electrostriction of a waveguide without considering pulse properties[37]. However, a theory about instantaneous interactions between a laser pulse and its excited phonons, to our knowledge, has not been reported yet.

The spectra of ultrafast pulses generated by mode-locked lasers have a frequency comb structure, although they generally cannot be used as precision frequency combs without stabilizing their carrier-envelope offset. In general, the frequencies of Stokes ($\omega - \Omega$) and anti-Stokes ($\omega + \Omega$) of forward SBS induced by a comb line ($\omega$) do not align with the comb lines making Stokes significantly stronger than anti-Stokes (Fig. 2). In this general case, the acoustic wave ($\Omega$) induced by different comb lines with different phases mainly cancel out each other as they have a bandwidth of about 1THz or larger, more than three orders of magnitude of $\Omega$, and thus the total forward SBS effect is so small that it is generally ignored in laser pulse generation and transmission. However, if the Stokes and anti-Stokes lines overlap with the comb lines, anti-Stokes can become stronger than Stokes at some moments; in other words, the acoustic wave is not only emitted but also absorbed by the comb lines at different pulse positions. The total forward SBS effect could be very small but the peak instantaneous effect will be significantly stronger than the general case, particularly at the resonant state where all the phonons at different frequencies generated or absorbed by the frequency comb are synchronized, resulting in the formation of a phonon pulse. This requires all the comb lines to be properly chirped in a dispersive media to adjust the phase difference among comb lines for the synchronization of phonons. We call this resonant state of forward SBS under phase-matching condition as SBSR. The frequency of phonon pulse caused by SBSR effect is $\Omega_R = N_R \cdot \delta\omega \approx \Omega$, where $\delta\omega$ is the mode separation of the frequency comb and $N_R$ is an integer (Fig. 2). A $\Omega_R$ frequency is covered by a forward SBS line peak at $\Omega$. SBSR effect depends on the total number of comb lines (pulse intensity and duration), meaning the resonant state is a few orders of magnitude stronger than forward SBS caused by an individual pair of comb lines. In this case, all comb lines link together by the generation and absorption of phonon pulses at any waveguide position and are synchronously modulated by the frequency of the phonon pulse $\Omega_R$, which enables passive frequency-modulation mode-locking and spectral profile self-regulation.

In a fiber laser, when SBS signals exceed the laser noise, forward SBS intrinsically synchronizes a lasing mode and the modes covered by its Stokes and anti-Stokes lines, causing random pulsing. Because any lasing mode could be the pump mode, eventually the lasing modes present in every pulse with a mode separation $\Omega_R$ will cover the full laser gain bandwidth only if their polarization evolutions in the cavity are nearly consistent and the laser is not intensity modulated. The pulses are constructed with different lasing modes ($N_R$ groups). They are down-chirped and exhibit nearly equal capabilities for emitting and absorbing phonons due to the SBSR effect (Supplemental Material and section 2.2 Adaptive soliton). The $\Omega_R$ phonons generated by different laser pulses will interfere with each other and be absorbed by the laser pulses, which modulate the phases of the laser pulses. If the SBSR effect is replaced by regular forward SBS, the capability of laser pulses to absorb phonons is very weak, and thus this phase modulation will not occur. The pulses with SBSR effect eventually will be synchronized via phonon generation and absorption, leading to passive frequency-modulation mode-locking.

In addition, SBSR effect will significantly suppress backward SBS in fibers because the comb lines down-chirped for the phase matching of SBSR at a forward SBS frequency ($\Omega_R$) will dramatically reduce their stimulation effect on backward Brillouin scattering with the frequency deviation from the resonance point ($\Omega_R$) about one to two orders of magnitude[38,39].

There is no clear way to directly measure the SBSR effect because it does not produce a signal different from the comb lines and the total acoustic wave generation or absorption is close to zero, but it can be indirectly observed as it will lead to several special properties to the laser pulses, including down-chirping the frequency comb and balancing TOD.

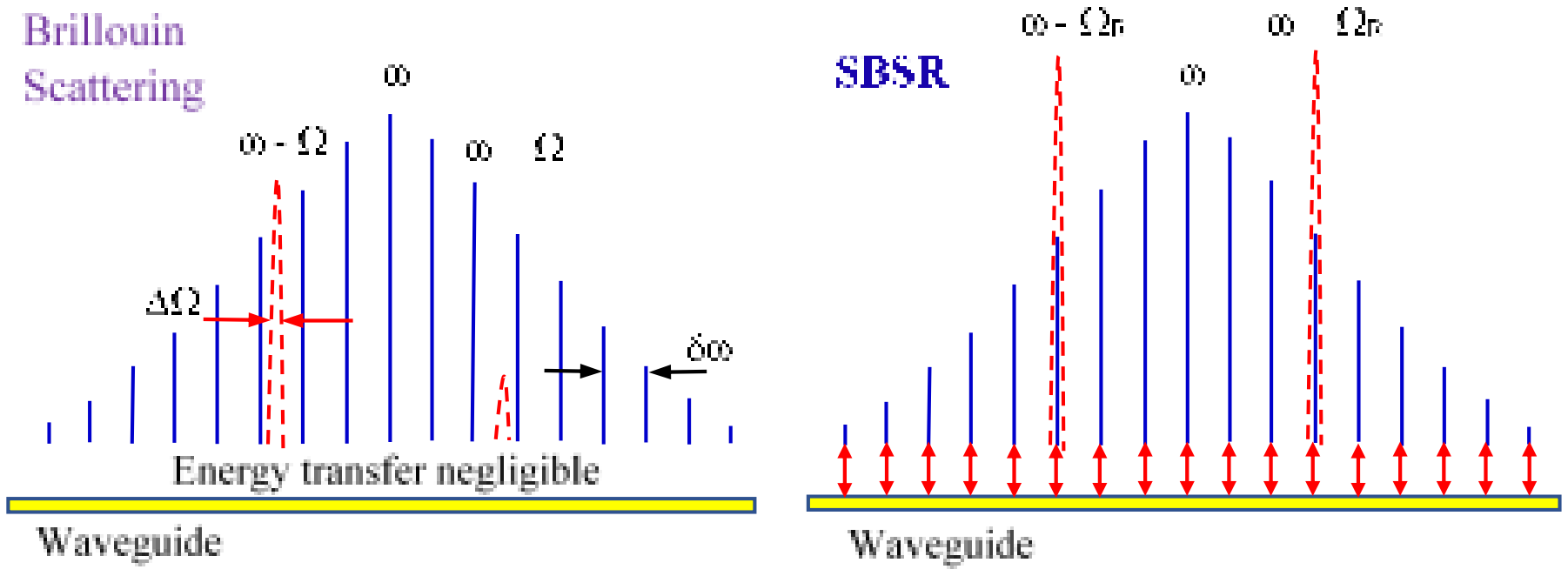

**Fig. 2 Schematic of the frequency condition for SBSR**

Generally, $\Omega \gg \delta\omega$. To satisfy the SBSR frequency condition, the laser should be designed or controlled such that its mode separation ($\delta\omega$) meets $N_R \cdot \delta\omega \approx \Omega$. The control of $\delta\omega$ has a tolerance which is estimated as the full width at -20 dB of the forward SBS lines. Within the tolerance, SBSR should be significantly stronger than backward SBS. It is about $10\Delta\Omega$, where $\Delta\Omega$ is the full width at half maximal (FWHM) of the forward SBS with Lorentzian line shape. The maximal $\Delta\Omega$ in single mode fibers is about 1 MHz[38] so this tolerance will be up to about 10 MHz. The figure depicts the effects of a single comb line, but this happens simultaneously for all comb lines. When SBSR occurs, significant energy transfer between the comb lines and the waveguide occurs, making all the comb lines modulated by the excited resonant acoustic frequency $\Omega_R = N_R \cdot \delta\omega$.

Most mode-locked lasers have a repetition rate larger than 10 MHz. Without cavity length control, the SBSR frequency condition is not readily achieved. More importantly, without pulse down-chirped to compensate the phase-mismatching caused by GVD and the Kerr effect, forward SBS induced by the frequency comb will not resonate and thus is negligible or noise-like in current mode-locked lasers. Some fiber lasers have mode separation of only a few MHz or even smaller than 0.1 MHz, making the SBSR frequency condition intrinsically satisfied, but their mode-locking is based on nonlinear polarization evolution (NPE) or saturable absorbers[36,40], which cannot down-chirp the laser pulse for SBSR effect achievement. Their mode-locking mechanism is fundamentally based on intensity modulation, where forward SBS of the intracavity fibers is not involved with the starting of mode-locking, although some harmonic mode-locking lasers use optomechanical interaction of laser pulses with a photonic crystal fiber to lock the harmonic repetition rate to the phonon frequency [35,36].

Utilizing Fourier transform and inverse Fourier transform, the nonlinear optical equation of SBSR effect can be derived (Supplemental Material) from the forward SBS coupled mode dynamic equations under phase matching condition[39,41] as,

$$\left(\frac{\partial}{\partial Z}+\frac{\alpha'}{2}\right)u(T,z) = -R_B\frac{\partial}{\partial T}\left(u(T,z)\frac{\partial}{\partial T}|u(T,z)|^2\right) \tag{1}$$

with the SBSR coefficient $R_B = C_B g_0/(2\omega_0\Omega_0)$, where $g_0$ is the power gain coefficient of forward SBS, $\omega_0$ and $\Omega_0$ are the center angular frequencies of the laser pulse and the acoustic wave, respectively. Parameter $C_B$ reflects the coupling ratio of a SBS line with its nearest laser mode. It is determined by the SBS line shape, with the maximal equal to 1 as their peaks overlap. $\alpha'$ is the laser power loss parameter along the propagation direction. Eq. (1) shows that the SBSR effect depends on the pulse intensity ($|u|^2$), chirp (encompassed within $\partial u/\partial T$), and duration (time differential of $u$ and $|u|^2$).

**2.2 Adaptive soliton**

By taking the SBSR effect into account, the pulse amplitude in a transmission waveguide obeys the nonlinear Schrödinger equation in the slowly varying envelope approximation[17-20,25-30] and can be described by

$$\frac{\partial}{\partial z}u = -R_B\frac{\partial}{\partial T}\left(u\frac{\partial}{\partial T}|u|^2\right) - i\frac{\beta_2}{2}\frac{\partial^2}{\partial T^2}u + \frac{\beta_3}{6}\frac{\partial^3}{\partial T^3}u + i\gamma|u|^2u + \frac{\Delta G}{2}u \tag{2}$$

The terms on the right-hand side of Eq. (2) are the actions of the following: the SBSR effect described by Eq. (1), GVD $\beta_2$, TOD $\beta_3$, Kerr nonlinearity $\gamma$, and the total power-independent gain ($\Delta G > 0$) or loss ($\Delta G < 0$) coefficient, respectively. $\Delta G$ includes the loss $\alpha'$ in Eq. (1). For simplicity, we first set $\beta_3 = 0$ to give the approximate solutions of adaptive solitons, and then we discuss what will happen when $\beta_3$ is taken into account. The ansatz of Eq. (2) with $\beta_3 = 0$ is a chirped soliton:

$$u(z,T) = U_0\exp[(\Delta G/2 - \sigma/2 + i\Delta k)z]\,\mathrm{sech}^{(1+i\rho)}\left(\frac{T}{\tau}\right)\exp\left(i\eta_2\tanh^2\left(\frac{T}{\tau}\right) + i\eta_4\tanh^4\left(\frac{T}{\tau}\right) + \cdots\right) \tag{3}$$

As $|\tanh(x)| < 1$, the high order chirp terms in Eq. (3) are negligible. Without TOD, all odd-order chirp terms ($\eta_{2n+1}$, $n \in \mathbb{Z}_{>0}$) are zero. Substituting Eq. (3) into Eq. (2), we obtain the parameters (Supplemental Material):

$$\sigma = -2\rho\beta_2/\tau^2 \tag{4}$$

$$\Delta k = (\rho^2 - 1)\beta_2/(2\tau^2) \tag{5}$$

$$\tau^2 = (\rho^2 - 2)\,\beta_2/(2P_0\gamma) \tag{6}$$

$$\rho = 0,\ 8R_BP_0/(3\beta_2) \tag{7}$$

$$\eta_2 = 3\rho/2 - 2R_BP_0/\beta_2 \tag{8}$$

with $P_0 = |u(0,Z)|^2$, the peak power of the soliton at position Z. As a sech$^2$-shaped pulse[42], its pulse duration is $1.763\tau$. The chirp parameter $\rho$ in Eq. (7) has two independent solutions, $\rho_1 = 0$ and $\rho_2 = 8R_BP_0/(3\beta_2)$. Accordingly, the power-dependent loss coefficient $\sigma$ (Eq. (4)) is $\sigma_1 = 0$ for $\rho = \rho_1$, and $\sigma_2 = -16R_BP_0/(3\tau^2)$ for $\rho = \rho_2$. It indicates the scale of energy transfer between solitons and the waveguide. $\sigma_2$ is negative for bright (positive $P_0$) solitons and positive for dark (negative $P_0$) solitons, corresponding to phonon absorption (waveguide cooling) and emission (waveguide heating), respectively. The wavenumber shift $\Delta k = P_0\gamma(\rho^2 - 1)/(\rho^2 - 2)$ is only caused by the Kerr effect.

The chirp parameters $\eta_2$ together with $\rho$ determine the spectrum profile of the adaptive soliton. For identification, we call $\eta_2$ as the regular main chirp parameter and $\rho$ as the adaptive chirp parameter which could be zero or nonzero and determines whether $\tau$ is SBSR effect dependent or not. The spectrum bandwidth (FWHM) of a chirped pulse $u(T) = \mathrm{sech}^{(1+i\xi)}(T/\tau)$ can be calculated as[42]:

$$\Delta\omega = 2\mathrm{arcosh}[2 + \cosh(\pi\xi)]/(\pi\tau). \tag{9}$$

Note that the chirp of a laser pulse is mainly determined by the second order chirp parameter, and $\mathrm{sech}^{i\xi}(T/\tau) = \exp\left[-i\xi\ln\left(\cosh(T/\tau)\right)\right]$. Comparing the $(T/\tau)^2$ term of the Taylor expansion of $-\xi\ln[\cosh(T/\tau)]$ with the phase of the solitons at $T = 0$, the bandwidth can be estimated from Eq. (9) with $\xi = \rho - 2\eta_2$, although their higher order chirps are different from that of $\mathrm{sech}^{(1+i\xi)}(T/\tau)$. The solitons are down-chirped with $\xi = -2\eta_2 = 4R_BP_0/\beta_2 < 0$ in anomalous waveguides and $\xi = -\rho/2 = -4R_BP_0/(3\beta_2) < 0$ in normal dispersion waveguides. It can be interpreted as that the soliton absorbs phonons at the leading edge and emits phonons at the trailing edge. For normal dispersion waveguides, the down-chirping effect is a special feature of the SBSR effect; it does not occur with any other known nonlinear optical effects.

The soliton chirps ($\xi$) are proportional to $R_BP_0$ (SBSR effect) and inversely proportional to $\beta_2$, but Kerr effect independent. Attempting to increase $P_0$ will increase the soliton chirp. On the other hand, increasing the chirp will reduce $P_0$ and the SBSR effect. This implies that the SBSR effect and the chirp of the soliton work as a feedback loop in pulse shaping. Therefore, this kind of soliton is super stable and adaptive to the waveguides meeting SBSR condition so we refer to them as adaptive solitons.

For anomalous dispersion, dark adaptive solitons will be generated if the average power is so large that $\rho_2^2 \geq 2$. To achieve bright adaptive solitons, Eq. (6) requires $\rho^2 < 2$. The solution is restricted to $\rho = \rho_1 = 0$ and $\sigma = 0$ which means it is non-dissipative if $\Delta G = 0$; otherwise, the soliton undergoes self-shortening and self-intensification, ultimately violating the requirement. Following the same area theorem of chirp-free solitons[25], $\tau^2P_0 = |\beta_2|/\gamma$, so its energy is low.

For normal dispersion, bright adaptive solitons can only be achieved when $\rho^2 > 2$ (Eq. (6)). This limits the solution to be $\rho = \rho_2 = \sqrt{2}P_0/P_{\mathrm{min}}$ with $P_{\mathrm{min}} = 3\sqrt{2}\beta_2/(8R_B)$. As $\rho^2 \gg 2$, the bandwidth ($\Delta\omega$) and the pulse duration ($1.763\tau$) are both proportional to $\sqrt{P_0}$. High pulse energy ($E \propto \tau^3$) and broadband spectrum can be simultaneously obtained from adaptive solitons without area theorem limitation.

The total transmission gain of an adaptive soliton includes power-independent loss $\alpha'$ and power-dependent loss $\sigma$. It can be expressed as $G_{as} = -\alpha' - \sigma$, where $\sigma$ is given by Eq. (4), and $\alpha'$ is an undetermined constant. Note that when the peak power $P_0$ approaches infinity, the photons absorbed or emitted by the waveguide in a limited area is negligible compared with the soliton energy. The total transmission gain or loss coefficient should approach zero for $\rho\to\infty$. This requires $\alpha' = -\sigma_\infty$. For the case of $\rho = \rho_1 = 0$, including bright solitons in anomalous dispersion waveguides and dark solitons in normal dispersion waveguides, $\sigma = 0$. Their $\alpha'$ and $G_{as}$ are zero. For the case of $\rho = \rho_2 = 8R_BP_0/(3\beta_2)$, including bright solitons in normal dispersion waveguides and dark solitons in anomalous dispersion waveguides, $\sigma = \sigma_\infty\rho^2/(\rho^2 - 2)$ with $\sigma_\infty = -3\beta_2\gamma/(2R_B)$. The total transmission gain (positive for normal dispersion) or loss (negative for anomalous dispersion) coefficients can be expressed as

$$G_{as} = 2\,G_2/(\rho^2 - 2) \tag{10}$$

with $G_2 = 3\beta_2\gamma/(2R_B)$ which can be calculated from the properties of the transmission waveguide. It can be designed by controlling the soliton energy or the properties of the delivery waveguide, including the GVD $\beta_2$, SBSR parameter $R_B$, and Kerr nonlinearity $\gamma$. Eq. (10) shows $G_{as}$ is adaptive, determined by ρ.

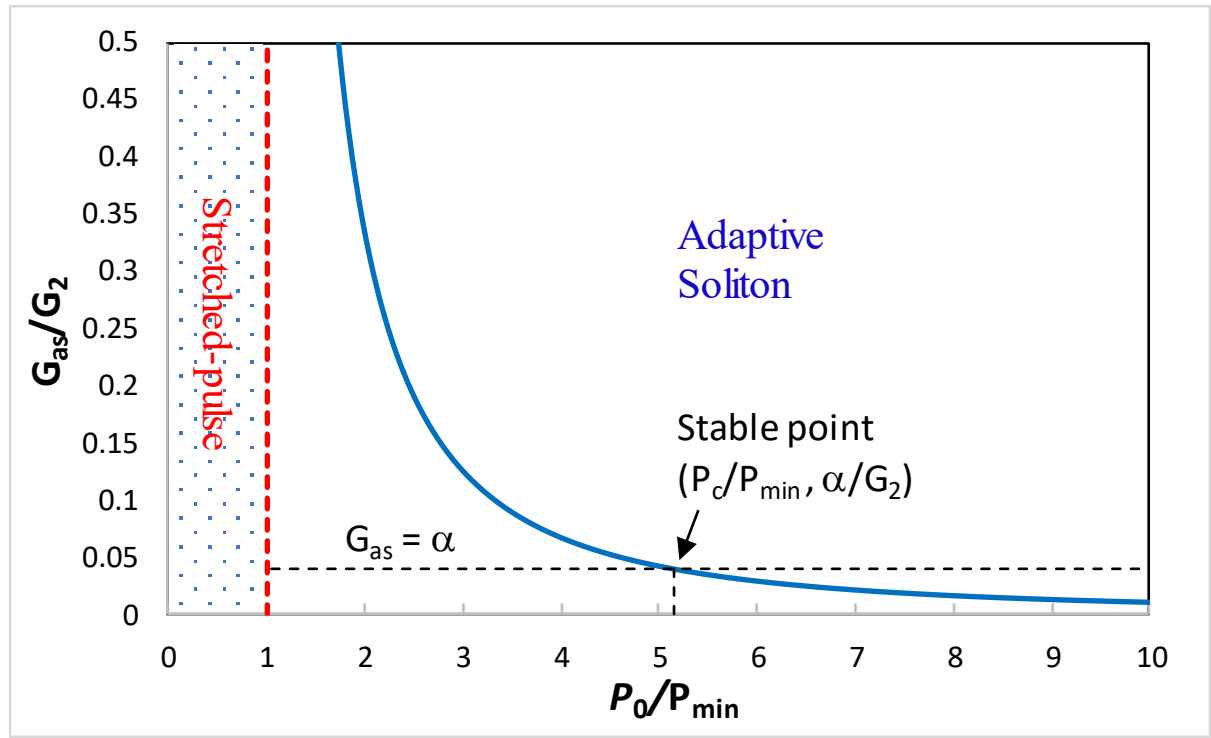

**Fig. 3 Theoretical adaptive transmission gain of adaptive solitons in normal dispersion waveguides.**

When $P_0$ is close to $P_{min}$ (or $\tau \rightarrow 0$), $G_{as}$ approaches infinity, preventing the adaptive soliton from transforming into a stretched-pulse during waveguide transmission. With the peak power increasing, $G_{as}$ will reduce until it is balanced by waveguide attenuation ($\alpha$), and then the soliton will be locked to a constant power $P_c$ due to the balance of gain and attenuation. Similarly, for $P_0 > P_c$, the waveguide attenuation will reduce the soliton power until it drops to $P_c$. For propagation in a long fiber, the output power of adaptive solitons is theorized to be $P_c$ no matter how the input soliton energy, the fiber length, and the fiber attenuation are.

The method for solving Eq. (2) with $\beta_3 = 0$ is still valid when we reintroduce TOD ($\beta_3 \neq 0$), provided that odd-order chirps are taken into account. It can be obtained that τ is determined by a quartic equation:

$$\tau^4 - \frac{\beta_2}{2P_0\gamma}[\rho^2 - 2]\tau^2 + \beta_3 f(\rho, \eta_2) = 0, \tag{6'}$$

where $f(\rho, \eta_2)$ is a function of the chirp parameters ρ and $\eta_2$. It is not difficult to figure out that each of the parameters τ, ρ, $\eta_2$, as well as $\eta_3$ should have four solutions for $\beta_3 \neq 0$, which degenerate to two (except for $\eta_3 = 0$) when $\beta_3 = 0$, as shown in Eqs. (3) to (8). However, it is complicated and tedious to obtain and solve the whole set of parametric equations so those are out of scope for this article. Fortunately, the transmission loss or gain σ is still described by Eq. (4), which is obtained from the zero order (T independent) parameter equation and that equation is independent of $\eta_n$ ($n \in \mathbb{Z}_{>0}$). We can reach two important conclusions without completely solving the parameters. First, the existence of solutions means that adaptive solitons persist when $\beta_3 \neq 0$. Second, from Eq. (4) we know that the sign of transmission loss or gain $\sigma$ is determined by the chirp parameter $\rho$ and GVD $\beta_2$. The balanced TOD $\beta_3$ only introduces a small variation to σ via τ. There is no extra loss item caused by $\beta_3$. This implies that dispersive radiation does not happen to adaptive solitons. No matter if $\beta_3$ is positive or negative, bright adaptive solitons are theorized to always have a transmission gain ($\sigma < 0$) in normal dispersion waveguides.

The above analysis indicates that the dispersions, including GVD and TOD, balanced by the effects of SBSR and SPM do not cause any loss to the solitons. This implies that no dispersive waves will be generated by adaptive solitons. Generally, dispersive waves in a laser cavity with large cavity dispersion will cause Kelly sidebands to the solitons if there are no intracavity elements, such as spectral filters or saturable absorbers, to suppress its enhancement under phase-matching conditions[25-27]. Note NPE which is often used in chirp-free soliton lasers is not strong enough to suppress Kelly-sidebands significantly. In an adaptive soliton laser, there is no intensity modulation element required and its cavity dispersion is large. Therefore, as long as TOD in the laser is not negligible, the absence of sidebands will indicate that TOD is balanced.

### 2.3 Adaptive soliton laser

To validate the theoretical predictions, an experimental adaptive soliton Yb-doped fiber laser[43] schematically shown in Fig. 4a was first constructed to test the feasibility of SBSR mode-locking, and then a portable prototype was built for measuring the pulse properties.

In the experimental setup, a ~35 meter HI1060 fiber is used to set the mode separation to about 5 MHz, so that the SBSR frequency condition can be intrinsically satisfied. The active fiber is a 1 meter single mode Yb doped fiber (Leikki YB1200-4/125). It is pumped by a 980nm laser diode via a WDM. A polarization sensitive isolator is used to make the lasing unidirectional, and combined a polarization controller (PC) to ensure the polarization and intensity of the laser is self-consistent in the cavity, which is a basic requirement for stable mode-locking achievement, different from NPE modulation. The laser is output from a coupler.

In this laser cavity, there are no saturable absorbers or spectrum filters. Therefore, the cubic-quintic Ginzburg–Landau equation in this laser is simplified as a regular nonlinear Schrödinger equation plus a term of constant gain or loss[28-30]. With the SBSR effect, the equation can be expressed as Eq. (2). The adaptive soliton in the cavity is basically the solution of Eq. (2), approximately described by Eqs. (4) to (8), if the GVD is not too small that TOD dominates the soliton chirp.

The laser first ran at CW at lower pump power, and then self-pulsing appeared when the intracavity power was larger than the threshold of forward SBS. By adjusting polarization control to maximize the self-pulsing peaks, the operation became Q-switched mode-locking (Fig. 4b and c). With the pump power increasing, the peak power and repetition rate of Q-switched mode-locking became higher and higher. After reaching a threshold, the SBSR effect overcame GVD and SPM, and the operation switched from Q-switched to SBSR mode-locking with a repetition rate of 5.26 MHz (Fig. 4b).

When the output power was about 29 mW, corresponding to pulse energy of 5.5 nJ, the spectrum of the soliton was very close to the typical emission spectrum of Yb-doped fibers (Fig. 4c). Its peak wavelength was 1035 nm and the bandwidth was about 12 nm, corresponding to TL 90 fs. It has no Kelly-sidebands or SPM peaks, distinguishing the laser pulse from chirp-free soliton and dissipative solitons generated from all-normal-dispersion fiber lasers[28-30]. Therefore, it can be preliminarily concluded that this laser pulse output from an all-normal dispersion laser does not belong to the same class as chirp-free solitons, dissipative solitons, or stretched pulse from a dispersion-managed laser; it is a new type of soliton, an adaptive soliton. It was generated from a laser with large dispersion and without intensity modulation elements. Note that the TOD of HI1060 fiber is comparable to that of SMF28 fiber, which is known to be non-negligible. The absence of Kelly-sidebands means that dispersive radiation was eliminated since TOD is balanced by the SBSR effect.

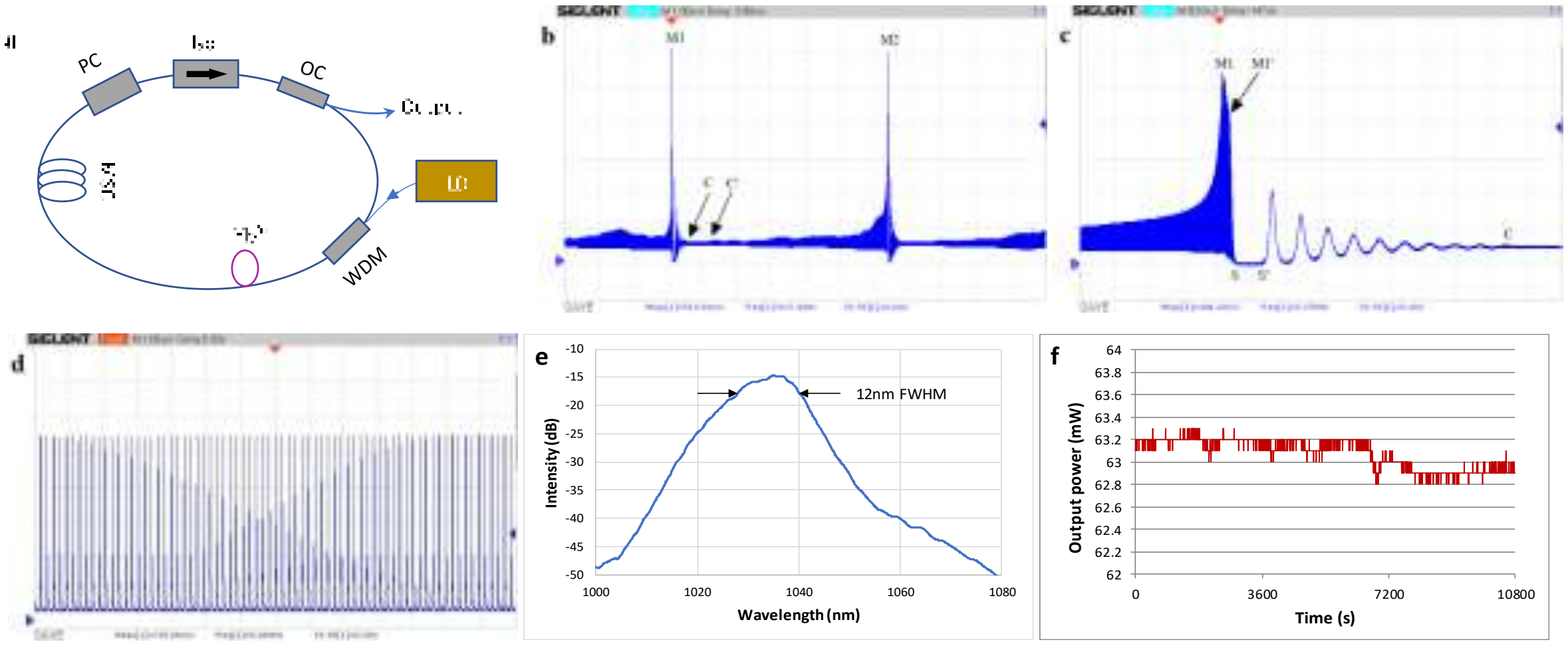

**Fig. 4 Experimental results of adaptive soliton laser**

(a**)** Schematic of the experimental adaptive soliton laser with the following components: $Yb^{3+}$, Yb doped fiber; MSC, mode separation control which is a HI1060 fiber; PC, polarization controller; Iso, polarization sensitive isolator; OC, output coupler; WDM, wavelength division multiplexer; LD, 976 nm laser diode. (b) Measured signals of Q-switched mode-locking at low power operation, time scale 1ms/div. The peaks M1 and M2 were the moments that the pulses were chirp-free. Each Q-switched mode-locking cycle consisted of multiple processes, including CW operation (from C to C'), random pulsing caused by forward SBS (from C'), down-chirped pulse generation via mode-locking, and pulse compression until the pulse became chirp-free (M2) due to up-chirp induced by positive GVD. (c) Zoom-in of the vicinity of one peak shows the detail of Q-switched mode-locking, time scale 50µs/div. During pulse compression, the SBSR effect decreased from its maximal to zero, while SPM increased to a maximal, accelerating the compression of the laser pulse. After the peak moment (M1), the laser pulse had no more SBSR effect and was stretched by positive GVD and SPM, while backward SBS surpassed forward SBS became dominant. The intensity variation from the peak was related to a period $T_0$ determined by backward SBS and GVD. It decreased slowly at the beginning, then dropped very rapidly when the elapsed time t was longer than $T_0/4$ (at M1'), and eventually reached zero at $t = T_0/2$ (S). The lasing halted due to backward SBS and the isolation introduced by the isolator. Meanwhile, a strong acoustic wave was generated in the fiber. With the acoustic wave phase changing and intensity damping, the lasing started again (at S') and then was Q-switched by the acoustic wave with a period of $T_0$ until it faded away (at C). After an uncertain short time CW operation, the next cycle of self-pulsing and mode-locking was excited again. (d**)** Pulse train under SBSR mode-locking, time scale 1µs/div, and (e) spectrum output from the experimental setup. (f) Output power of the portable prototype for three hours after warm-up.

In the portable prototype, the polarization controller was constructed with waveplates. The misalignment tolerance of each individual waveplate is larger than ±10°, more than 5 times of that for NPE mode-locking[45]. Without any temperature control, it is turn-key mode-locking in the environment temperature range from 19 C° to 23 C°. Also, it is insensitive to vibrations as repeated impacts to the exterior casing did not disrupt mode-locking operation. The high stability of the adaptive soliton laser is also shown in Fig. 4d. The output power fluctuation is smaller than ±0.5%. Its output power is 63 mW and has a repetition rate of 5.118 MHz, giving pulse energy of 12.3 nJ.

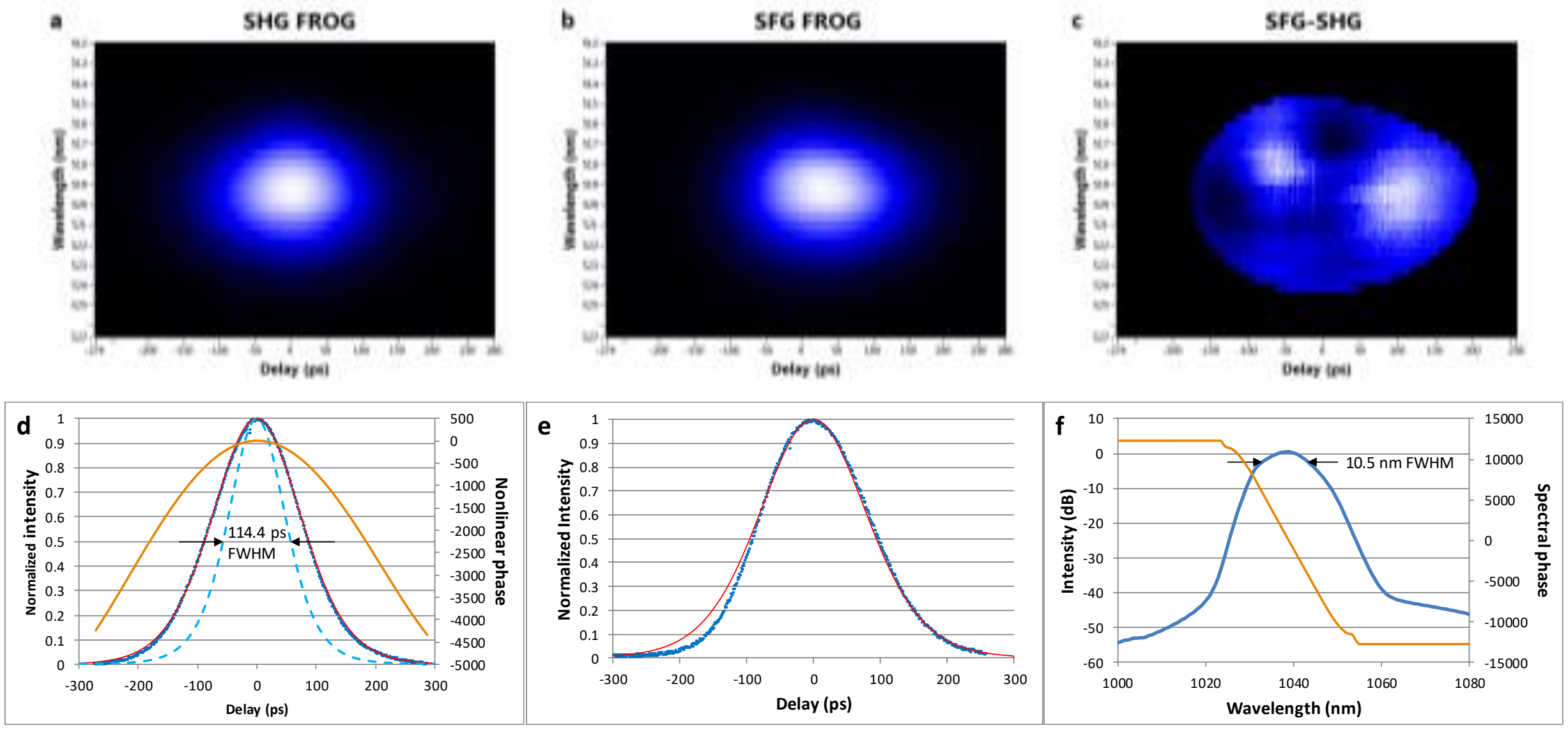

**Fig. 5 Properties of adaptive solitons output from the portable prototype measured by FROG**

(a) Normalized SHG FROG, (b) normalized SFG FROG with zero delay point shifted by a bandpass filter, and (c) the SFG FROG subtracted by SHG FROG with a 30 ps delay. The data acquisition zero-time delay setting of SFG FROG was the same as that of SHG FROG, but its actual zero-time delay of the peak signal was delayed a few picoseconds when a bandpass filter was inserted in the fixed optical path of the autocorrelator. This is helpful for the determination of the leading edge and the trailing edge of the measured pulse in the data. (d) The autocorrelation curve obtained from SHG FROG (blue dot), the fitting autocorrelation (red solid line) of $\text{sech}^2$-shaped pulse with duration 114.4 ps (cyan dashed line), and the nonlinear phase of the soliton retrieved from the SHG FROG (orange solid line), (e) the cross-correlation curve obtained from SFG FROG (blue dot) and the fitting of autocorrelation of $\text{sech}^2$-shaped pulse with duration 122.5 ps (red solid line), and (f) the spectrum of the soliton measured by an optical spectrum analyzer (blue solid line), and the spectral phase retrieved from the SHG FROG (orange solid line).

The theory predicts that an adaptive soliton is down-chirped. This is impossible for a laser pulse generated from an all-normal-dispersion fiber laser based on current mode-locking technologies. A Second Harmonic Generation (SHG) Frequency Resolved Optical Gating (FROG) is often used for pulse chirp measurement. The standard approach for determining the chirp sign is by inserting a glass plate into the input path to cause a visible chirp change. However, it becomes impracticable for a pulse chirped significantly longer than 10ps, due to the required dispersion being too large. To determine the chirp sign, we modified a SHG FROG whose signal is always symmetrical around the zero delay (Fig. 5a) as a sum frequency generation (SFG) FROG to break the symmetrical pattern, which is realized by inserting a bandpass filter (FGB25 from Thorlabs) ~45° in the fixed optical path of the autocorrelator of the SHG FROG. The filter significantly reduced the intensity of wavelength components longer than 1039nm, the peak wavelength of the soliton, which stretches the soliton according to Fourier transform. The cross-correlation of two laser pulses with different pulse durations features some properties of the longer pulse. Compared with SHG FROG in Fig. 5a, the SFG FROG in Fig. 5b shows a longer cross-correlation pulse and some asymmetrical features. To make the asymmetrical features more obvious, we subtract SHG FROG from SFG FROG at different delays until bright signals show up at both sides of negative delay and positive delay for identifying the presence of up-chirp or down-chirp, as shown in Fig. 5c. This figure clearly shows that the spectrum of the stretched soliton is shortened by the filter. The leading edge of the stretched pulse, which corresponds to the shorter delay, has a higher frequency, meaning the soliton is down chirped.

The theoretical prediction of the down-chirped soliton being $sech^2$-shaped is proven by the fitting of the autocorrelation curve obtained from the SHG FROG (Fig. 5d) by subtracting the white noise. From the fitting, a pulse duration ($\Delta T$) of 114.4 ps is obtained. The nonlinear phase of the laser pulse is retrieved from the SHG FROG data as $\Phi_{NL} = \pi(-0.0245T^2 + 1.2 \cdot 10^{-6}T^3 + 9 \cdot 10^{-8}T^4)$ with time unit (T) in picoseconds, plotted in Fig. 5d. The pulse duration broadening is mainly induced by the linear chirp with a chirp rate $C = -0.0245$ THz/ps, corresponding to a second order chirp parameter $\xi = 2\pi C(\Delta T/1.763)^2 = -644$. The third order chirp appears for adaptive solitons only when TOD is not negligible and comes from the balance between TOD and the SBSR effect. It causes the asymmetry of the spectrum with respect to the peak frequency to keep the pulse $sech^2$-shaped. The fourth order chirp parameter leads to a variation of the spectral slope.

The cross-correlation curve obtained from SFG FROG is shown in Fig. 5e. Its leading edge is closer to the original shape (Fig. 5d), and its trailing edge is approximately fitted by the autocorrelation of $sech^2$-shape pulse with duration of 122.5 ps. This is because the filter reduces more energy of lower frequency at the trailing edge of the down-chirped soliton. Fig. 5f shows the spectrum output from the laser. The bandwidth is about 10.5 nm. The spectral phase retrieved from the SHG FROG is also shown in Fig. 5f. The chirp of an adaptive soliton is very close to linear, indicated by the linear spectral phase shown in Fig. 5f and the domination of the linear chirp on $\Phi_{NL}$, which means the soliton completely compressible. The spectrum shown in Fig. 5f has no Kelly sidebands or SPM peaks, same as the experimental result shown in Fig. 4c, meaning TOD is balanced by the SBSR effect.

The chirp parameter $\rho = -2\xi$ is calculated as 1288. Note that the nonlinear index $n_2$ of regular fused silica fibers is $2.2 \times 10^{-20}$ $m^2/W$, corresponding to $\gamma = 4.7 km^{-1}W^{-1}$ of HI1060 fiber at 1039nm[46]. From Eq. (6) and the measured pulse duration, we can calculate that the peak power of the soliton in the laser cavity is 0.88 kW, about nine times of the laser output. Accordingly, the SBSR coefficient $R_B$ of HI1060 fiber @1039nm can be calculated from Eq. (7) as 11.5 $ps^2/(km \cdot W)$ .

Forward Brillouin scattering in a regular optical fiber has many modes between 200-600 MHz, with the maximal power gain coefficient $g_0$ about 10 $m^{-1}W^{-1}$ for single mode bare fibers in air[38]. The effective modes that meet SBSR frequency condition could be one or multiple of them. Normally, higher frequency modes have broader linewidths, making them easier to meet the SBSR frequency condition. Assuming the center acoustic frequency of SBSR is 500 MHz and $C_B = 1$, we can calculate the effective $g_0$ from $R_B$ which is defined in Eq. (1) as about 1.3 $m^{-1}W^{-1}$, a moderate value compared with that of the forward Brillouin scattering modes.

The transmission gain parameter $G_2$ which is defined in Eq. (10) can be calculated as 12.8 $km^{-1}$. From Eq. (10), we know it is large enough for the transmission gain to balance the attenuation of the HI1060 fiber (~1.5dB/km) if $\rho$ is smaller than 7.

## 3. Discussion

The process of passive frequency-modulation mode-locking achievement is well interpreted by the SBSR effect. The measured soliton properties which include smooth spectrum, down-chirp, high mode-locking stability, and $sech^2$ pulse shape, agree with the theoretical predictions. From the measurement results of the pulse duration, bandwidth, and chirp rate, the power gain coefficient of forward SBS is calculated as a value that is within expectations. All these imply the SBSR effect exists and the theory is valid. Adaptive solitons based on the SBSR effect are expected to break the limitations of dissipative solitons in generation, transmission, and amplification.

The SBSR effect couples all comb lines together, enabling passive frequency-modulation mode-locking for adaptive soliton generation. This new mode-locking technology was demonstrated being significantly stronger than NPE mode-locking which is the most effective mode-locking approach for dissipative soliton lasers achieving ultrashort laser pulses[47] and more stable than Kerr-lens mode-locking[24,48,49] commonly used in solid-state lasers. Compared with current ultrashort laser pulses, adaptive solitons with passive loop control constructed by the SBSR effect, GVD, and the Kerr effect are significantly more stable, possess high pulse contrast, and have stronger self-healing capability. They are adaptively down-chirped which breaks soliton area theorem limitation on pulse energy in normal dispersion lasers. They are more reliable and consistent, and thus have higher manipulability in applications, such as modulating the spectrum and chirp for qubit manipulation[8]. With above advantages, SBSR mode-locking will become the most popular mode-locking approach for ultrashort pulse generation.

In this paper, low repetition rate lasers with mode separation intrinsically meeting SBSR frequency condition are used to generate adaptive solitons because they are simple, stable and relatively low cost. To achieve high repetition rate (>100 MHz) adaptive solitons, either the laser cavity length has to be controlled such that the mode separation meets the SBSR frequency condition (up to the maximal forward Brillouin frequency in the laser), or by utilizing regular high repetition rate approaches, such as harmonic mode-locking[35] or repetition rate multiplication[50], to obtain high repetition rate laser pulses with a small mode separation (<10 MHz) that intrinsically meets the SBSR frequency condition. Their stability might not be as good as low repetition rate SBSR mode-locked lasers, but with fundamental modes locked by SBSR effect, they should still be significantly more stable than the lasers of the same repetition rate with fundamental modes locked by NPE technology, as well as its extended approaches, including nonlinear optical loop mirror and nonlinear amplifying loop mirror.

The SBSR effect has the unique capabilities of balancing TOD to eliminate sidebands and dispersive radiation, and down-chirping laser pulses in transmission media to balance the up-chirp effect of SPM and normal GVD, which enables the formation of adaptive solitons in normal dispersion waveguides. From the theoretical predictions, the transmission gain of adaptive solitons in normal dispersion waveguides can be expected, although it has not been experimentally demonstrated yet. This effect is negligible for larger energy (>1 nJ) adaptive solitons, but can be significant for small energy ones. As shown in Fig. 3, $G_{as}$ will trap adaptive solitons to a state that balances the gain and the waveguide attenuation, and becomes lossless in long transmission waveguides. It has the potential to enable repeaterless long-haul soliton communications which was proposed in 1973[51] and initially demonstrated in 1980[52], but is not commercially successful partly due to the propagation loss of chirp-free solitons being insufficiently low and the self-healing capability being too low to withstand soliton interactions[11]. Repeaterless long-haul soliton communication would remove the need for frequent signal amplifiers and the reliance on external electricity, which simplifies fiber deployments, eliminates Gordon-Haus jitter[53], improves reliability, and reduces the operational power footprint in an ever-expanding network. Based on adaptive solitons, all normal dispersion fibers could be used for soliton communication, and the communication bands could be extended from near IR to visible light.

Current CPA fiber amplifiers are obstructed by gain narrowing effect from obtaining TL <200 fs pulses[54] and high-energy similaritons. Similaritons (parabolic pulses) were considered as some of the highest-quality ultrafast pulses generated in optical fibers. However, their required bandwidth is proportional to the cube root of the pulse energy[55], making it very difficult to achieve energies greater than 10 μJ. Adaptive solitons with adaptive chirp and spectrum self-regulation could break through gain narrowing effect limitations. This would enable high energy ultrashort pulse amplifications without the need of nonlinear spectral broadening[54] which has stability and pulse quality concerns, or a gain medium with super broad gain spectrum, such as Ti:sapphire crystal that only works for a specific wavelength. Their chirp can be controlled by designing the dispersion, as well as the Brillouin scattering parameters, of the transmission fibers. Using adaptive solitons as the seeds, fiber CPA would become simple and effective for high energy ultrashort pulse achievement. They will speed up the developments of ultrashort pulses in cold ablation and other high-power applications[3-6,12].

With high stability, high contrast, moderate manipulability, and low transmission loss, adaptive solitons are the ideal sources for most ultrafast applications, in particular for those with fiber delivery. They are poised to revolutionize ultrafast laser systems, providing a reliable, distortion-free platform for diverse fields and to accelerate industrial applications of ultrashort laser pulses.

## Supplementary Material

This section includes the detailed derivations of the nonlinear optical equation of the SBSR effect Eq. (1), and the procedures of solving the NLSE involving the SBSR effect.

***Derivation of nonlinear optical equation of the SBSR effect***

In a waveguide, the forward SBS interaction among an acoustic wave $b(\Omega,z)$, and the E-fields of two CW laser lines $a_0(\omega,z)$ and $a_1(\omega-\Omega,z)$, respectively, can be described at the phase matching condition by the coupled mode equations[39,41].

$$\left(v_0^{-1}\frac{\partial}{\partial t}+\frac{\partial}{\partial z}+\frac{\alpha_0'}{2}\right)a_0(\omega,z)=-i\omega\frac{Q^*}{p_0}a_1(\omega-\Omega,z)b(\Omega,z), \tag{11}$$

$$\left(v_1^{-1}\frac{\partial}{\partial t}+\frac{\partial}{\partial z}+\frac{\alpha_1'}{2}\right)a_1(\omega-\Omega,z)=-i(\omega-\Omega)\frac{Q}{p_1}a_0(\omega,z)b^*(\Omega,z), \tag{12}$$

$$\left(\frac{\partial}{\partial t}+v_b\frac{\partial}{\partial z}+\Gamma_b\right)b(\Omega,z)=-i\Omega\frac{Q}{e_b}a_0(\omega,z)a_1^*(\omega-\Omega,z). \tag{13}$$

where Ω, $\omega$, and $\omega-\Omega$ are the angular frequencies of the acoustic wave and the two CW lasers, respectively. $\upsilon_0$, $\upsilon_1$, and $\upsilon_b$ are the group velocities of the two CW lasers and the acoustic wave, respectively. $p_0$ and $p_1$ are the power normalization factors of the two lasers, respectively. $\alpha'$ is the laser power loss parameter along the direction z. Q is the opto-mechanical perturbation overlap, $e_b$ is the phonon energy, $\Gamma_b$ is the acoustic decay parameter, and the $*$ denotes complex conjugation. The laser line $a_0(\omega,z)$ is the anti-Stokes of the laser line $a_1(\omega-\Omega,z)$, and $a_1(\omega-\Omega,z)$ is the Stokes of $a_0(\omega,z)$. Eqs. (11) and (12) describe their dynamic interaction, with the strengths depending on the phase-matched acoustic wave $b(\Omega,z)$. Eq (13) describes the acoustic wave $b(\Omega,z)$ generated or absorbed by the pair of laser lines $a_0(\omega,z)$ and $a_1(\omega-\Omega,z)$.

For the acoustic wave generated by a broadband laser, the pump and the Stokes or anti-Stokes have the same spectrum and the same phase ($a_0(\omega,z)=a(\omega,z)$ and $a_1(\omega-\Omega,z)=a(\omega-\Omega,z)$) under SBSR frequency condition and phase matching conditions. In this case, each frequency of the acoustic wave is generated by all the frequencies of the broadband laser, and then Eq. (13) can be rewritten with the transformations $T_b=t-z/\upsilon_b$ and Z = z as

$$\left(v_b\frac{\partial}{\partial Z}+\Gamma_b\right)b(\Omega,Z)=-i\Omega\frac{Q}{e_b}\int_{-\infty}^{\infty}a(\omega,Z)a^*(\omega-\Omega,Z)\,d\omega. \tag{13'}$$

Performing inverse Fourier transform on Eq. (13'), the equation becomes

$$\left(v_b \frac{\partial}{\partial Z} + \Gamma_b\right) B(T_b, Z) = \int_{-\infty}^{\infty} \left[-i\Omega \frac{Q}{e_b} \int_{-\infty}^{\infty} a(\omega, Z) a^*(\omega - \Omega, Z)\, d\omega\right] \exp(i\Omega T_b)\, d\Omega \qquad (14)$$

where $B(T_b, Z) = \int_{-\infty}^{\infty} b(\Omega, Z) \exp(i\Omega T_b)\, d\Omega$. Note that only the term $\exp(i\Omega T_b)$ is a function of $T_b$ and $\partial[\exp(i\Omega T_b)]/\partial T_b = i\Omega \exp(i\Omega T_b)$. Eq. (14) can be rewritten as

$$\left(v_b \frac{\partial}{\partial Z} + \Gamma_b\right) B(T_b, Z) = -\frac{Q}{e_b} \frac{\partial}{\partial T_b} \int_{-\infty}^{\infty} \int_{-\infty}^{\infty} a(\omega, Z) a^*(\omega - \Omega, Z) \exp(i\Omega T_b) d\omega d\Omega \quad (15)$$

Note that $\exp(i\Omega T_b) = \exp(i\omega T_b) \exp[i(\Omega - \omega) T_b]$. Switching the integration order of ω and Ω and moving Ω independent terms out of the integration of Ω, Eq. (15) can be further rewritten as

$$\left(v_b \frac{\partial}{\partial Z} + \Gamma_b\right) B(T_b, Z) = -\frac{Q}{e_b} \frac{\partial}{\partial T_b} \int_{-\infty}^{\infty} a(\omega, Z) \exp(i\omega T_b) \left[\int_{-\infty}^{\infty} a^*(\Omega', Z) [exp(i\Omega' T_b)]^* d\Omega\right] d\omega \quad (16)$$

with $\Omega' = \omega - \Omega$. Note $u^*(T_b, Z) = \int_{-\infty}^{\infty} a^*(\Omega', Z) [exp(i\Omega' T_b)]^*\, d\Omega'$ and $d\Omega' = -d\Omega$ in this integration, Eq. (16) can be expressed as

$$\left(v_b \frac{\partial}{\partial Z} + \Gamma_b\right) B(T_b, Z) = \frac{Q}{e_b} \frac{\partial}{\partial T_b} \int_{-\infty}^{\infty} a(\omega, Z) \exp(i\omega T_b) u^*(T_b, Z)\, d\omega \qquad (17)$$

As a function of $T_b$, $u^*(T_b, Z)$ is independent of ω, and thus can be moved out of the integration of ω in Eq. (17). Eventually, the total acoustic wave as a function of time can be obtained from Eq. (17) as

$$\left(v_b \frac{\partial}{\partial Z} + \Gamma_b\right) B(T_b, Z) = \frac{Q}{e_b} \frac{\partial}{\partial T_b} |u(T_b, Z)|^2, \qquad (18)$$

where $B(T_b, Z)$ and $u(T_b, Z)$ are the inverse Fourier transforms of $b(\Omega, Z)$ and $a(\omega, Z)$, respectively. For slowly varying envelope approximation, the laser pulse remains approximately the same over the length scale of the acoustic decay (< 0.1 mm in an optical fiber), therefore the first item at the left-hand side of Eq. (18) is negligible. The acoustic wave can be calculated as

$$B(t, z) \approx \frac{Q}{\Gamma_b e_b} \frac{\partial}{\partial T_b} |u(T_b, Z)|^2 = \frac{Q}{\Gamma_b e_b} \frac{\partial}{\partial t} |u(t, z)|^2. \qquad (19)$$

The amplitude of the acoustic wave is proportional to the slope of the laser power. When a laser pulse passes through a position, the excited acoustic wave starts at zero and ends up at zero, which means the total energy change of the laser pulse caused by acoustic wave alone is close to zero. At either the leading or trailing edge of a single-peak laser pulse, the laser first emits and then absorbs the acoustic wave. Accordingly, the phase of the laser pulse is modulated by the acoustic wave. Since most of laser pulse energy is within the FWHM bandwidth, the laser pulse absorbs more than it emits phonons at the leading edge and the opposite way at the trailing edge, which makes the laser pulse down-chirped.

Meanwhile, each frequency of the broadband laser will be scattered by all the frequencies of the acoustic wave, such that Eq. (11) and (12) can be rewritten, with the transformation $T = t - z/\upsilon$ and Z = z, as

$$\left(\frac{\partial}{\partial Z} + \frac{\alpha'}{2}\right) a(\omega, Z) = -i\omega \frac{Q^*}{p} \int_{-\infty}^{\infty} a(\omega - \Omega, Z) b(\Omega, Z) d\Omega \qquad (11')$$

$$\left(\frac{\partial}{\partial Z} + \frac{\alpha'}{2}\right) a(\omega, Z) = -i\omega \frac{Q}{p} \int_{-\infty}^{\infty} a(\omega + \Omega, Z) b^*(\Omega, Z)\, d\Omega \qquad (12')$$

Performing inverse Fourier transform on Eq. (11') or (12'), which is similar to the derivation from Eq. (13') to Eq. (18), and substituting the acoustic wave with Eq. (19), we can obtain the nonlinear optical equation of Eq. (1) with the maximal SBSR effect,

$$\left(\frac{\partial}{\partial Z}+\frac{\alpha'}{2}\right)u(T,Z) = -R_{B0}\frac{\partial}{\partial T}\left(u(T,Z)\frac{\partial}{\partial T}|u(T,Z)|^2\right)$$

with the SBSR coefficient $R_{B0} = g_0/(2\omega_0\Omega_0)$, where $g_0 = 2\omega_0\Omega_0|Q|^2/(p^2\Gamma_b e_b)$ is the power gain coefficient of Brillouin scattering in units of $m^{-1}W^{-1}$, and $\omega_0$ and $\Omega_0$ are the center frequencies of the laser pulse and the acoustic wave, respectively. Here, $|u(T,Z)|^2$ is already normalized with power unit W.

***Procedure of solving the NLSE involving SBSR effect***

The ansatz of the NLSE Eq. (2) is a chirped soliton given by Eq. (3),

$$u(z,T) = U_0\exp[(\Delta G/2-\sigma/2+i\Delta k)Z]\mathrm{sech}^{(1+i\rho)}\left(\frac{T}{\tau}\right)\exp\left(i\eta_2\tanh^2\left(\frac{T}{\tau}\right)+i\eta_4\tanh^4\left(\frac{T}{\tau}\right)+\cdots\right)$$

The basic parameters and functions used to solve the simultaneous equations of the parameters in Eq. (3) are:

The power of the pulse:

$$|u|^2 = {U_0}^2\exp\big((\Delta G-\sigma)Z\big)\mathrm{sech}^2\left(\frac{T}{\tau}\right) = P_0\mathrm{sech}^2\left(\frac{T}{\tau}\right),$$

where $P_0 = U_0^2\exp[(\Delta G-\sigma)Z]$ is the peak power of soliton at position Z.

The first partial derivative of $|u|^2$ to T:

$$\frac{\partial}{\partial T}|u|^2 = -\frac{2P_0}{\tau}\tanh\left(\frac{T}{\tau}\right)\mathrm{sech}^2\left(\frac{T}{\tau}\right)$$

The second partial derivative of $|u|^2$ to T:

$$\frac{\partial^2}{\partial T^2}|u|^2 = \frac{2P_0}{\tau^2}\mathrm{sech}^2\left(\frac{T}{\tau}\right)\left(-1+3\tanh^2\left(\frac{T}{\tau}\right)\right)$$

The first partial derivative of u to T:

$$\frac{\partial}{\partial T}u = \frac{u}{\tau}\left[-(1+i\rho)\tanh\left(\frac{T}{\tau}\right)+i2\eta_2\tanh\left(\frac{T}{\tau}\right)\mathrm{sech}^2\left(\frac{T}{\tau}\right)+i4\eta_4\tanh^3\left(\frac{T}{\tau}\right)\mathrm{sech}^2\left(\frac{T}{\tau}\right)\right]$$

The second partial derivative of u to T, omitting the higher orders ($n \geq 4$) of $\tanh^n\left(\frac{T}{\tau}\right)$:

$$\frac{\partial^2}{\partial T^2}u = \frac{u}{\tau^2}\Big\{(i\rho+1)^2 + \mathrm{sech}^2\left(\frac{T}{\tau}\right)\Big[\left(\rho^2-2-i3\rho+i2\eta_2\right)+\left(4\rho\eta_2-4\eta_2^2-i10\eta_2+i12\eta_4\right)\tanh^2\left(\frac{T}{\tau}\right)\Big]\Big\}$$

To obtain the simultaneous equations of the parameters in the ansatz of the NLSE including the SBSR effect, the field $u$(T, z) in Eq. (2) has to be substituted with Eq. (3). To do so, all the terms in Eq. (2) are to be calculated one by one. The higher orders ($n \geq 4$) of $\tanh^n\left(\frac{T}{\tau}\right)$ are ignored.

The term at the left-hand side of Eq. (2) can be calculated as:

$$\frac{\partial}{\partial z}u = (\Delta G/2-\sigma/2+i\Delta k)u \qquad (20)$$

The first term at the right-hand side of Eq. (2) can be calculated as:

$$-R_B\frac{\partial}{\partial T}\left(u\frac{\partial}{\partial T}|u|^2\right)=\frac{2R_BP_0}{\tau^2}\mathrm{sech}^2\left(\frac{T}{\tau}\right)\left[1-\left(4+\mathrm{i}\rho-i2\eta_2\right)\tanh^2\left(\frac{T}{\tau}\right)\right]u \tag{21}$$

The second term at the right-hand side of Eq. (2) can be calculated as:

$$-i\frac{\beta_2}{2}\frac{\partial^2}{\partial T^2}u=-i\frac{\beta_2}{2\tau^2}u\left\{(\mathrm{i}\rho+1)^2+\mathrm{sech}^2\left(\frac{\mathrm{T}}{\tau}\right)\left[\left(\rho^2-2-i3\rho+i2\eta_2\right)+\left(4\eta_2\rho-4\eta_2{}^2-i10\eta_2+i12\eta_4\right)\tanh^2\left(\frac{\mathrm{T}}{\tau}\right)\right]\right\} \tag{22}$$

The third term at the right-hand side of Eq. (2) can be calculated as:

$$\mathrm{i}\gamma|u|^2u=\mathrm{i}\gamma P_0\mathrm{sech}^2\left(\frac{T}{\tau}\right)u \tag{23}$$

Substituting Eqs. (20) to (23) into Eq. (2), the right-hand side of Eq. (2) can be expressed as $\left[K_0+\mathrm{sech}^2\left(\frac{\mathrm{T}}{\tau}\right)\sum C_{2n}\tanh^{2n}\left(\frac{T}{\tau}\right)\right]u$, with n = 0, 1, 2, … . Except for the first term $K_0u$ equal to the left-hand side of Eq. (2), all other parameters $C_{2n}$ should be equal to zero to make Eq. (2) valid. Now we have the simultaneous equations of the parameters:

For the $K_0$ term, the parameter equation is:

$$-\sigma/2+i\Delta k=-i\frac{\beta_2}{2\tau^2}(\mathrm{i}\rho+1)^2 \tag{24}$$

It has no terms with the chirp parameters $\eta_n$. The real part of Eq. (24) is Eq. (4). The imaginary part of Eq. (24) is Eq. (5).

For the $C_0$ term, the parameter equation is:

$$\frac{2R_BP_0}{\tau^2}-i\frac{\beta_2}{2\tau^2}\left(\rho^2-2-i3\rho+i2\eta_2\right)+i\gamma P_0=0 \tag{25}$$

From the imaginary part of Eq. (25), we obtain Eq. (6). The real part of Eq. (25) is Eq. (8).

For the $C_2$ term, the parameter equation is:

$$-2R_BP_0\left(4+\mathrm{i}\rho-i2\eta_2\right)-i\frac{\beta_2}{2}\left(4\eta_2\rho-4\eta_2{}^2-i10\eta_2+i12\eta_4\right)=0 \tag{26}$$

The imaginary part of Eq. (26) gives:

$$-R_BP_0\left(\rho-2\eta_2\right)-\beta_2\eta_2\left(\rho-\eta_2\right)=0 \tag{27}$$

Substituting Eq. (8) into Eq. (27), we obtain a parabola equation of ρ. Its solutions are given in Eq. (7). The real part of Eq. (26) is

$$\eta_4=4R_BP_0/\left(3\beta_2\right)+5\eta_2/6 \tag{28}$$

The chirp parameter $\eta_4$ can be calculated from Eq. (28).

**Data availability:** All data supporting the findings of this study are available in the article.

# References

1. C. Xu, and F. W. Wise, Recent advances in fiber lasers for nonlinear microscopy. *Nature Photonics*, **7**, 875-882 (2013).

2. C. L. Hoy, O. Ferhanoğlu, M. Yildirim, K. H. Kim, S. S. Karajanagi, K. M. C. Chan, J. B. Kobler, S. M. Zeitels, and A. Ben-Yakar, Clinical Ultrafast Laser Surgery: Recent Advances and Future Directions. *IEEE J. Sel. Top. Quantum Electron.*, **20**, 242-255 (2013).
3. F. H. Loesel, J. P. Fischer, M. H. Götz, C. Horvath, T. Juhasz, F. Noack, N. Suhm, and J. F. Bille, J. F. Non-thermal ablation of neural tissue with femtosecond laser pulses. *Appl. Phys. B*, **66**, 121–128 (1998).
4. M. Malinauskas, A. Žukauskas, S. Hasegawa, Y. Hayasaki, V. Mizeikis, R. Buividas, and S. Juodkazis, S. Ultrafast laser processing of materials: from science to industry. *Light: Science & Applications*, **5**, e16133-e16133 (2016).
5. S. K. Saha, D. Wang, V. H. Nguyen, Y. Chang, J. S. Oakdale1, and S.-C. Chen, Scalable submicrometer additive manufacturing. *Science*, **366**, 105-109 (2019).
6. X. Xu, T. Wang, P. Chen, C. Zhao, J. Ma, D. Wei, H. Wang, B. Niu, X. Fang, D. Wu, S. Zhu, M. Gu, M. Xiao, and Y. Zhang, Femtosecond laser writing of lithium niobate ferroelectric nanodomains. *Nature,* **609**, 496-501 (2022).
7. D. Meshulach, and Y. Siberberg, Coherent quantum control of two-photon transitions by a femtosecond laser pulse. *Nature,* **396**, 239-242 (1998).
8. R. de Vivie-Riedle, and U. Troppmann, U. Femtosecond Lasers for Quantum Information Technology. Chem. Rev., **107**, 5082-5100 (2007).
9. T. D. Ladd, F. Jelezko, R. Laflamme, Y. Nakamura, C. Monroe, and J. L. O'Brien, Quantum computers. *Nature*, **464**, 45-53(2010).
10. K. Jhuria, V. Ivanov, D. Polley, Y. Zhiyenbayev, W. Liu, A. Persaud, W. Redjem, W. Qarony, P. Parajuli, Q. Ji, and A. J. Gonsalves, Programmable quantum emitter formation in silicon. *Nature Communications*, **15**, 4497 (2024).
11. A. Hasegawa, Optical soliton: Review of its discovery and applications in ultra-high-speed communications. *Frontiers in Physics*, **10**, 1044845 (2022).
12. T. Ditmire, J. Zweiback, V. P. Yannovsky, T. E. Cowan, G. Hays, and K. B. Wharton, Nuclear fusion from explosions of femtosecond laser-heated deuterium clusters. *Nature*, **398**, 489-492 (1999).
13. A. M. Weiner, Ultrafast optical pulse shaping: A tutorial review. *Optics Communications,* **284**, 3669-3692 (2011).
14. F. Verluise, V. Laude, Z. Cheng, Ch. Spielmann, and P. Tournois, Amplitude and phase control of ultrashort pulse by use of an acousto-optic programmable dispersive filter: pulse compression and shaping. *Opt. Lett.,* **25**, 575-577 (2000).
15. A. C. Scott, F. Y. Chu, and D. W. McLaughlin, The soliton: a new concept in applied science. *Proceedings of the IEEE,* **61**, 1443-1483 (1973).
16. N. J. Zabusky, and M. D. Kruskal, Interaction of "solitons" in a collisionless plasma and the recurrence of initial states. *Physical Review Letters*, **15**, 240 (1965).
17. P. Grelu, and N. Akhmediev, Dissipative solitons for mode-locked lasers. *Nat. Photonics*, **6,** 84-92(2012).

18. P. K. A. Wai, C. R. Menyuk, Y. C. Lee, and H. H. Chen, Nonlinear pulse propagation in the neighborhood of the zero-dispersion wavelength of monomode optical fibers. *Opt. Lett.* **11**, 464-466(1986).

19. N. Akhmediev, and M. Karlsson, Cherenkov radiation emitted by solitons in optical fibers. *Phys. Rev. A*, Vol. 51, 2602-2607 (1995).

20. T. I. Lakoba, G. P. Agrawal, Effects of third-order dispersion on dispersion-managed solitons. *J. Opt. Soc. Am. B*, Vol. 16, 1332-1342 (1999).

21. E. Picholle, C. Montes, C. Leycuras, O. Legrand, and J. Botineau, Observation of dissipative superluminous solitons in a Brillouin fiber ring laser. *Phys. Rev. Lett.* **66,** 1454–1457 (1991).

22. E. V. Vanin, A. I. Korytin, A. M. Sergeev, D. Anderson, M. Lisak, and L. Vázquez, Dissipative optical solitons. *Phys. Rev. A* **49,** 2806–2811 (1994).

23. K. Tamura, E. P. Ippen, H. A. Haus L. and E. Nelson, 77-fs pulse generation from a stretched-pulse mode-locked all-fiber ring laser. *Opt. Lett.*, **18**, 1080-1082 (1993).

24. U. Keller, Ultrafast solid-state laser oscillators: a success story for the last 20 years with no end in sight. *Applied Physics B,* **100**, 15-28 (2010).

25. H. A. Haus, Mode-Locking of Lasers. *IEEE J. Sel. Top. Quantum Electron*., **6**, 1173-1185 (2000).

26. A. F. J. Runge, D. D. Hudson, K. K. K. Tam, C. M. de Sterke, and A. Blanco-Redondo, The pure-quartic soliton laser. *Nat. Photonics,* **14,** 492–497 (2020).

27. S. M. J. Kelly, Characteristic sideband instability of periodically amplified average soliton. *Electron. Lett.*, **28**, 806–807 (1992).

28. W. H. Renninger, A. Chong, and F. W. Wise, Dissipative solitons in normal-dispersion fiber lasers. *Physical Review A–Atomic, Molecular, and Optical Physics*, **77**, 023814 (2008).

29. W. H. Renninger, A. Chong, and F. W. Wise, Area theorem and energy quantization for dissipative optical solitons. *J. Opt. Soc. Am. B*, **27**, 1978–1982 (2010).

30. B. Oktem, C. Ülgüdür, and F. Ö. Llday, Soliton-similariton fiber laser. *Nat. Photonics*, **4**, 307-311 (2010).

31. S. Xu, A. Turnali, and M. Y. Sander, Group-velocity-locked vector solitons and dissipative solitons in a single fiber laser with net-anomalous dispersion. *Scientific Reports*, **12**, 6841 (2022).

32. J. P. Gordon, Theory of the soliton self-frequency shift. *Opt. Lett*., **11**, 662-664 (1986).

33. F. M. Mitschke, and L. F. Mollenauer, Discovery of the soliton self-frequency shift. *Opt. Lett*., **11**, 659-661 (1986).

34. Y. Bai, M. Zhang, Q. Shi, S. Ding, Y. Qin, Z. Xie, X. Jiang, and M. Xiao, Brillouin-Kerr soliton frequency combs in an optical microresonator. *Physical Review Letters* **126**, 063901 (2021).

35. M. Pang, X. Jiang, W. He, G. K. L. Wong, Onishchukov, G., Joly, N. Y., Ahmed, G., Menyuk, C. R., and Russell, P. ST. J. Stable subpicosecond soliton fiber laser passively mode-locked by gigahertz acoustic resonance in photonic crystal fiber core. *Optica* **2**, 339-342 (2015).

36. W. He, M. Pang, C. R. Menyuk, P. St. J. Russell, Sub-100-fs 1.87 GHz mode-locked fiber laser using stretched-soliton effects. *Optica* **3**, 1366-1372 (2016).

37. M. S. Kang, A. Nazarkin, A. Brenn, P. St. J. Russell, Tightly trapped acoustic phonons in photonic crystal fibres as highly nonlinear artificial Raman oscillators, *Nature Physics* **5**, 276-280 (2009).

38. E. Layosh, E. Zehavi, A. Bernstein, M. Hen, M. Holsblat, O. Pearl, and A. Zadok, Forward Brillouin Scattering in few-mode fibers. *Light: Science & Applications, **14,** 242 (2025)*.

39. P. St. J. Russell, D. Culverhouse, and F. Farahi, Theory of Forward Stimulated Brillouin Scattering in Dual-Mode Single-Core Fibers, IEEE J. Quantum Electron, 27, 836-842 (1991).

40. S. Kobtsev, S. Kukarin, and Y. Fedotov, Ultra-low repetition rate mode-locked fiber laser with high-energy pulses. Optics Express, ***26***, 21936-21941 (2008).

41. C. Wolff, M. J. A. Smith, B. Stiller, and C. G. Poulton, Brillouin scattering—theory and experiment: tutorial. *JOSA B*, **38**, 1243-1269 (2021).

42. P. Lazaridis, G. Debarge, and P. Gallion, Time–bandwidth product of chirped sech2 pulses: application to phase–amplitude-coupling factor measurement. *Opt. Lett.*, **20**, 1160-1162 (1995).

43. Y. Chen, Adaptive Soliton Laser. PCT patent application WO/2026/015166 (2026).

44. S. Song, A. Jung, and K. Oh, High-temperature sensitivity in stimulated Brillouin scattering of 1060 nm single-mode fibers. *Sensors*, ***19***, 4731 (2019).

45. M. E. Fermann, M. J. Andrejco, Y. Silberberg, and M. L. Stock, Passive mode locking by using nonlinear polarization evolution in a polarization-maintaining erbium-doped fiber. *Opt. Lett*., **18**, 894-896 (1998).

46. P. Kabaciński, T. M. Kardaś, Y. Stepanenko, and C. Radzewicz, Nonlinear refractive index measurement by SPM-induced phase regression. *Opt. Express*, **27**, 11018-11028 (2019).

47. W. Fu, L. G. Wright, P. Sidorenko, S. Backus, and F. W. Wise, Several new directions for ultrafast fiber lasers [Invited]. *Opt. Express*, **26**, 9432-9463 (2018).

48. F. Salin, J. Squier, and M. Piché, Mode locking of Ti:Al2O3 lasers and self-focusing: a Gaussian approximation. *Opt. Lett*. **16**, 1674-1677 (1991).

49. Y. C. Chen, and W. Z. Lin, Thick lens model for self-focusing in Kerr medium. *Appl. Phys. Lett*., **73**, 429-431 (1998).

50. S. Min, Y. Zhao, and S. Fleming, Repetition rate multiplication in figure-eight fibre laser with 3 dB couplers, *Optics Communications* **277**, 411-413 (2007).

51. A. Hasegawa, and F. Tappert, Transmission of stationary nonlinear optical pulses in dispersive dielectric fibers. I. anomalous dispersion. *Appl. Phys. Lett*., **23**, 142-144 (1973).

52. L. F. Mollenauer, R. H. Stolen, and J. P. Gordon, Experimental observation of picosecond pulse narrowing and solitons in optical fibers. *Phys. Rev. Lett*., **45**, 1095 –1098 (1980).

53. J. P. Gordon, and H. A. Haus, Random walk of coherently amplified solitons in optical fiber transmission. *Opt. Lett.*, **11**, 665-667 (1986).

54. P. Sidorenko, W. Fu, F. Wise, Nonlinear ultrafast fiber amplifiers beyond the gain-narrowing limit. *Optica*, **6**, 1328-1333 (2019).

55. V. I. kruglov, B. C. Thomsen, J. M. Dudley, and J. D. Harvey, Self-similar propagation and amplification of parabolic pulses in optical fibers, *Phys. Rev. Lett.*, 84, 6010 (2000).

## Acknowledgements

The laser development was self-funded by Lasoliton LLC. Thanks to Prof. Xi-Cheng Zhang, the Institute of Optics, University of Rochester, Rochester, NY 14627 for his arrangement of the FROG measurement to demonstrate the key properties of adaptive solitons. Thanks to grants NSF 2020249 and NSF 2152081 which supported the author, Jiacheng Zhao in pursuing his doctoral degree. Also we would like to thank Mr. Jinhui Yang and Mr. Wensheng Yang for their help in the preparation of experimental materials.

**Funding**

Self-funded by Lasoliton LLC

**Author's contributions:**

Y. C. discovered the SBSR effect, performed mathematical derivations, designed and built the lasers, and drafted the paper;

J. Z. designed and built the FROG setup, and measured the pulse duration and chirp;

Q. Z. designed the FROG layout.

B. C. built components for the portable laser and revised the paper;

**Competing interests:** Y.C. and B.C. are involved in commercializing the adaptive soliton laser at Lasoliton LLC. The methods and laser described in this paper are patent pending.